\documentstyle[12pt,epsfig]{article}
\newcommand{\simlt}  {\raisebox{-.6ex}{$\stackrel{\textstyle <}{\sim}$}}

\begin{document}
\begin{flushright}
RAL-TR-2009-008 \\
18 Oct 2009 \\
\end{flushright}
\vspace{2 mm}
\begin{center}
{\Large
The Matrix of Unitarity Triangle Angles for Quarks}
\end{center}
\vspace{3mm}
\begin{center}
{P.\ F.\ Harrison\footnotemark[1] \\
Department of Physics, University of Warwick,\\
Coventry, CV4 7AL. UK.}
\end{center}
\begin{center}
{and}
\end{center}
\begin{center}
{S.\ Dallison\footnotemark[2] and W.\ G.\ Scott\footnotemark[3] \\
Rutherford Appleton Laboratory,\\
Chilton, Didcot, Oxon OX11 0QX. UK.}
\end{center}
\vspace{3mm}
\begin{abstract}
\baselineskip 0.6cm
\noindent
In the context of quark (as for lepton) mixing,
we introduce the concept of the matrix
of unitarity triangle angles $\Phi$,
emphasising that it carries
equivalent information to the complex mixing matrix $V$ itself.
The angle matrix $\Phi$ has the added advantage, 
with respect to $V$, 
of being both basis- and phase-convention independent
and consequently observable
(indeed several $\Phi$-matrix entries, 
eg.\ $\Phi_{cs}=\alpha$, $\Phi_{us}=\beta$ etc.\
are already long-studied
as directly measurable/measured
in $B$-physics experiments). 
We give complete translation formulae between 
the mixing-matrix and angle-matrix representations.
In terms of Wolfenstein parameters, 
the invariant flavour-symmetric condition ${\rm Det} \, K=0$
(consistent with both quark and lepton data)
predicts $\cos \Phi_{cs}=\bar{\eta}\lambda^2$. 
We go on to consider briefly the present state 
of the experimental data on the full angle matrix
and some of the prospects for the future,
with reference to both the quark and lepton cases.  
\end{abstract}
\begin{center}
\end{center}
\footnotetext[1]{E-mail:p.f.harrison@warwick.ac.uk}
\footnotetext[2]{deceased}
\footnotetext[3]{E-mail:w.g.scott@rl.ac.uk}
\newpage
\baselineskip 0.6cm

\noindent {\bf 1. Introduction, Concept and Motivation}
\vspace{2mm}

\noindent
Following the  
pioneering early papers \cite{bigi} 
on $CP$ violation in $B$-meson decays,
and the tremendous successes of the $B$-factory experiments 
(see e.g.\ \cite{belle}), 
understanding of $CP$ violation
within the standard three-generation (CKM \cite{ckm}) scenario
has continued to grow.
Nonetheless,
on the key role 
of the unitarity triangles in the phenomenology,
the specific contribution of Aleksan et al.\ \cite{aleksan} 
and others \cite{lebed}
in effectively parameterising the CKM matrix in terms of 
four unitarity-triangle angles,
still merits further emphasis and development. 
In the present paper, 
building on the above \cite{aleksan} \cite{lebed},
we introduce
``the matrix of unitarity triangle angles for the quarks'', $\Phi$,
as a useful conceptual (and notational) advance
in the study of quark mixing (as for the leptons \cite{bhs}).
Our angle matrix ($\Phi$) closely mirrors,
and is in fact entirely equivalent to 
the complex mixing matrix $V$ itself (the CKM matrix \cite{ckm}),
but with the important advantage 
of being at once both real and basis- and phase-convention independent. 
Section~2 gives the explicit proofs of equivalence.
Section~3 gives 
some relevant Wolfenstein expansions\cite{wolf}. 

Concerning the complex mixing matrix $V$, 
it has long been appreciated \cite{branco}
that essentially all mixing observables
(including the magnitudes of the $CP$-violating asymmetries)  
are determined by any four independent modulii
- up to, in fact, only an overall sign ambiguity 
affecting all $CP$-violating asymmetries together in a correlated way.
Indeed, it is with this last proviso in mind, 
that we say that the matrix of moduli:
\begin{eqnarray} 
    \matrix{  \hspace{0.3cm} d \hspace{0.2cm}
               & \hspace{0.3cm} s \hspace{0.2cm}
               & \hspace{0.4cm} b  \hspace{0.2cm} }
                                      \hspace{0.8cm}
\hspace{1.8cm}
    \matrix{  \hspace{0.3cm} d \hspace{0.4cm}
               & \hspace{0.3cm} s \hspace{0.4cm}
               & \hspace{0.4cm} b  \hspace{0.4cm} }
                                      \hspace{1.1cm} \nonumber \\\
|V|=
\matrix{ u \hspace{0.01cm} \cr
         c \hspace{0.01cm} \cr
         t \hspace{0.01cm} }
\left(\matrix{
|V_{ud}| & |V_{us}| & |V_{ub}| \vspace{5pt} \cr
|V_{cd}| & |V_{cs}| & |V_{cb}| \vspace{4pt} \cr
|V_{td}| & |V_{ts}| & |V_{tb}| \cr
} \right)
\hspace{0.35cm}
\simeq
\hspace{0.35cm}
\matrix{ u \hspace{0.01cm} \cr
         c \hspace{0.01cm} \cr
         t \hspace{0.01cm} }
\left(\matrix{
0.974 & 0.226 & 0.004 \cr
0.226 & 0.973 & 0.042 \cr
0.009 & 0.041 & 0.999 \cr
} \right) \hspace{0.80cm}
\label{modckm}
\end{eqnarray}
is {\em essentially}\/ equivalent
to the complex mixing matrix itself 
(the numerical values of the moduli \cite{pdg} 
are displayed in Eq.~\ref{modckm} 
without their experimental errors, for simplicity). 
To facilitate a comparison,
we immediately introduce and display, on a similar footing,
the matrix of unitarity triangle (UT) angles for the quarks:
\begin{eqnarray}
    \matrix{  \hspace{0.2cm} d \hspace{0.2cm}
               & \hspace{0.2cm} s \hspace{0.2cm}
               & \hspace{0.2cm} b  \hspace{0.2cm} }
                                      \hspace{0.9cm}
\hspace{1.6cm}
    \matrix{  \hspace{0.2cm} d \hspace{0.2cm}
               & \hspace{0.2cm} s \hspace{0.3cm}
               & \hspace{0.2cm} b  \hspace{0.3cm} }
                                      \hspace{1.2cm} \nonumber \\\
\Phi=
\matrix{ u \hspace{0.01cm} \cr
         c \hspace{0.01cm} \cr
         t \hspace{0.01cm} }
\left(\matrix{
\Phi_{ud} & \Phi_{us} & \Phi_{ub} \vspace{5pt} \cr
\Phi_{cd} & \Phi_{cs} & \Phi_{cb} \vspace{4pt} \cr
\Phi_{td} & \Phi_{ts} & \Phi_{tb} \cr
} \right)
\hspace{0.35cm}
\simeq
\hspace{0.35cm}
\matrix{ u \hspace{0.01cm} \cr
         c \hspace{0.01cm} \cr
         t \hspace{0.01cm} }
\left(\matrix{
1^o & 22^o &  157^o \cr
67^o & 90^o & 23^o \cr
112^o & 68^o   & \sim \! 0^o \cr
} \right) \hspace{0.80cm}
\label{Phi=degrees}
\end{eqnarray}
where the entries are positive internal angles.
The row and column sums 
of the angle matrix (Eq.~\ref{Phi=degrees}) are all $180^o$,
while of course the moduli (squared)
in any row or column of the moduli matrix (Eq.~\ref{modckm}) sum to unity. 
Note further that the labelling of the rows and columns
is identical between the two matrices (Eq.~\ref{modckm} vs. Eq.~\ref{Phi=degrees}), 
and we see that the angle matrix already starts to ``mirror'' the mixing matrix,
as advertised above.

The mixing matrix elements, $V_{\alpha i}$,
represent directly the amplitudes for transition
between the up-type flavours (mass eigenstates) $u$, $c$, $t$ in the rows
and the down-type flavours  (mass eigenstates) $d$, $s$, $b$ in the columns.
We define the angle matrix entries (Eq.~\ref{Phi=degrees}) 
as the phases of the corresponding plaquette products $\Pi_{\alpha i}$ \cite{plaq} (see Section 2):
\begin{eqnarray}
    \matrix{  \hspace{3.2cm} d \hspace{1.0cm}
               & \hspace{1.0cm} s \hspace{1.0cm}
               & \hspace{1.0cm} b  \hspace{1.3cm} } \nonumber \\
\Phi:=
\matrix{ u \hspace{0.01cm} \cr
         c \hspace{0.01cm} \cr
         t \hspace{0.01cm} }
\left(\matrix{
{\rm Arg} \; (-\Pi_{ud}^*) & {\rm Arg} \; (-\Pi_{us}^*) & {\rm Arg} \; (-\Pi_{ub}^*) \vspace{5pt} \cr
{\rm Arg} \; (-\Pi_{cd}^*) & {\rm Arg} \; (-\Pi_{cs}^*) & {\rm Arg} \; (-\Pi_{cb}^*) \vspace{4pt} \cr
{\rm Arg} \; (-\Pi_{td}^*) & {\rm Arg} \; (-\Pi_{ts}^*) & {\rm Arg} \; (-\Pi_{tb}^*) \cr
} \right).
\label{Phi=ArgPi}
\end{eqnarray}
The minus sign and the complex conjugation are needed 
in Eq.~\ref{Phi=ArgPi} to convert from external to internal angles, 
maintaining consistency with existing conventions.

To appreciate the 
labelling of the angle matrix 
(Eqs.~\ref{Phi=degrees}-\ref{Phi=ArgPi})
we recall that any unitarity triangle
is defined by the inner product
of two given rows (or columns) of the complex mixing matrix $V$.
So here unitarity triangles are simply indexed
by the single (row or column) flavour label
{\em not} featuring in the inner product.
\begin{figure}[h]
\epsfig{figure=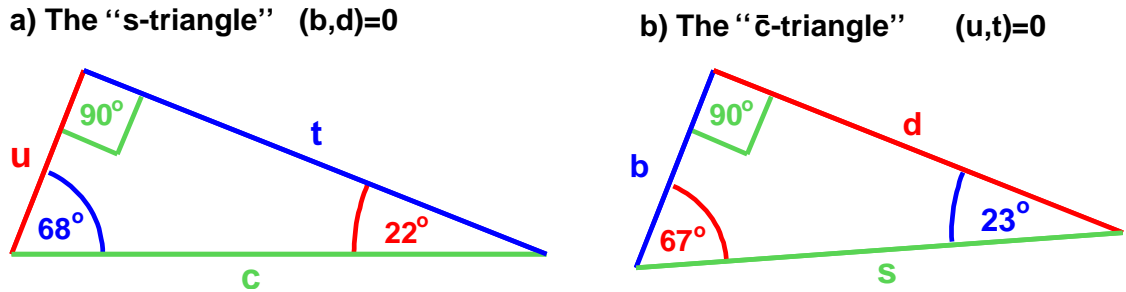,width=15.8cm}
\caption{ The indexing of unitarity triangles and their angles: 
a) The familiar row-based 
$(b,d)=:s$-triangle
and b) the closely similar 
$(u,t)=:\bar{c}$-triangle.
The indexing of the angles
follows from the flavour sub-amplitude {\em opposite} the angle (see text).}
\label{fig:sol}
\end{figure}
Then, with any inner product comprising the sum 
of three sub-amplitudes, each corresponding
to a particular intermediate quark flavour (mass eigenstate)
and to a particular side of the triangle, 
we have that any angle within a given triangle is simply labelled 
by the flavour label associated with the side {\em opposite} to that angle (see Figure~1).

It should now be clear that each row and each column 
of the angle matrix
corresponds to a particular unitarity triangle,
whereby there are three row-based triangles 
(labelled $u$, $c$, $t$, respectively)
and three column-based triangles
(labelled $d$, $s$, $b$, respectively)
giving, as is well-known, six unitarity triangles in all \footnote{
The complex conjugate triangles $\bar{d}$, $\bar{s}$ $\dots$ etc.,
formed reversing the order of arguments in the inner products
(e.g.\ the $(u,t)=(t,u)^*=:\bar{c}$-triangle, see Figure~1b),
are {\em not} counted separately here.}.
Each row/column then lists the angles of the corresponding triangle 
in (mass) order.
Clearly we have that 
each angle $\Phi_{\alpha i}$ appears 
in just one row-based triangle 
and just one column-based triangle,
as immediately 
specified by its row and column indices respectively.

Given the inherent difficulty 
of visualising six unitarity triangles at once,
each sharing each angle
with just one of the other triangles,
we expect the $\Phi$-matrix,
as the experimental focus (see Section~4)
moves on after the $B$-factory era,  
soon to come to be seen as the natural way 
to appreciate at a glance
all the standard-model (SM) weak-phases 
and their inter-relations, 
i.e.\ angles in common, 
relevant $180^o$ sums etc.
Importantly, if incidentally, it also offers 
a simple and definitive naming convention
for the (SM) quark UT angles,
to parallel that for the leptons~\cite{bhs},
free of the arbtrariness of prior nomenclatures (see Section~4)
arising in the case of the quarks. \\

\noindent {\bf 2. The Equivalence of the Mixing Matrix and the Angle Matrix}
\vspace{3mm}

\noindent
The definition of the $\Phi$-matrix 
given in the previous section (Eq.~\ref{Phi=ArgPi}) 
relies on that of the plaquette products.
Any given plaquette product $\Pi_{\alpha i}$
is obtained from the mixing matrix $V$,
by deleting the row and column 
containing the element $V_{\alpha i}$
to leave the complementary $2 \times 2$ sub-matrix, or ``plaquette'' \cite{plaq}.
The plaquette product $\Pi_{\alpha i}$ is then formed
by multiplying the four elements of the plaquette together,
with the appropriate pair of diagonally-related 
elements complex conjugated:
\begin{equation}
\Pi_{\alpha i}:=V_{\beta j} V_{\beta k}^* V_{\gamma k} V_{\gamma j}^*
\label{plaquette}
\end{equation}
where the indices are cyclically defined, 
i.e.\ in the columns, 
$i,j,k$ $(i \ne j \ne k)$
retain the cyclic order of $d,s,b$
and similarly for the up-type quarks in the rows
(equivalently, in terms of ``generation number'' we may write: $j=i+1$ mod 3, $k=i+2$ mod 3,~etc.).

Such plaquette products $\Pi_{\alpha i}$
constitute the minimal non-trivial 
loop amplitudes possible in flavour space \cite{ven09},
and are ubiquitous as interference terms
in e.g. squared penguin amplitudes,
after summing over the intermediate flavour in the loop
(particularly in the leptonic case, 
neutrino oscillations etc. \cite{pkq},
plaquette products have also been referred to as ``boxes''\cite{weil}).
Thus $\Phi$ is seen as a ``loop-space variable'' \cite{loops}, 
relating to the complex mixing matrix $V$ 
much in the way that, e.g., the observable field-strength tensor $F$
relates to the vector potential $A$ in electromagnetism. 


The matrix of plaquette products then takes the explicit form:
\begin{eqnarray}
    \matrix{  \hspace{2.2cm} d \hspace{1.1cm}
               & \hspace{1.1cm} s \hspace{1.1cm}
               & \hspace{1.1cm} b  \hspace{1.3cm} }  \nonumber \\
\Pi=
\matrix{ u \hspace{0.01cm} \cr
         c \hspace{0.01cm} \cr
         t \hspace{0.01cm} }
\left( \matrix{ 
  V_{tb} V_{ts}^* V_{cs} V_{cb}^*
                      & V_{td} V_{tb}^* V_{cb} V_{cd}^*
                                 & V_{ts} V_{td}^* V_{cd} V_{cs}^*  \cr
  V_{ub} V_{us}^* V_{ts} V_{tb}^*
                      & V_{ud} V_{ub}^* V_{tb} V_{td}^*
                                    & V_{us} V_{ud}^* V_{td} V_{ts}^*   \cr
  V_{cb} V_{cs}^* V_{us} V_{ub}^*
                      & V_{cd} V_{cb}^* V_{ub} V_{ud}^*
\label{Pi}                                    & V_{cs} V_{cd}^* V_{ud} V_{us}^*} \right)
\end{eqnarray}
where the pattern of complex conjugation
is seen to follow 
in straightforward analogy to the usual pattern
of computation of a $3 \times 3$ determinant
in terms of cofactors.
Likewise 
the labelling of the $\Pi$-matrix entries
is analogous to the labelling of the entries in 
``the matrix of cofactors'',
as encountered, e.g.\ in calculating a matrix inverse. 

The plaquette products 
are basis- and phase-convention independent
complex numbers.
As is well known, 
all the imaginary parts are equal \cite{cecj}: 
\begin{equation}
-\Pi^*=:
\left( \matrix{ K_{ud} & K_{us} & K_{ub} \cr
            K_{cd} & K_{cs} & K_{cb} \cr
            K_{td} & K_{ts} & K_{tb} } \right)
\hspace{2mm} + \hspace{2mm}
i \, \left( \matrix{ J & J & J \cr
                J & J & J \cr
                J & J & J } \right) \label{PiKJ}
\label{kplusj}
\end{equation}
and furthermore define $J$, the Jarlskog $CP$ invariant \cite{cecj}.
We remark that alternating signs ($\pm J$) often encountered
for the imaginary parts of the plaquette products 
do not enter here, with the cyclic definitions specified above.

To establish an equivalence
between the mixing matrix and the angle matrix
we still have to show that 
we can re-obtain the mixing matrix
starting from the angle matrix.
We shall take it as given \cite{branco}
that the matrix of mixing moduli
is (essentially) equivalent to the complex mixing matrix,
and content ourselves in the first instance 
with showing how to obtain the mixing-matrix moduli $|V_{\alpha i}|$,
starting from the angles.

We begin by defining, 
in terms of the $\Phi$-matrix,
a ${\rm Sin} \, \Phi$ matrix: 
\begin{equation}
{\rm Sin} \, \Phi:=
\left( \matrix{  \sin \Phi_{ud} & \sin \Phi_{us} & \sin \Phi_{ub} \cr
            \sin \Phi_{cd} &  \sin \Phi_{cs} &  \sin \Phi_{cb} \cr
             \sin \Phi_{td} &  \sin \Phi_{ts} &  \sin \Phi_{tb} } \right)
=
J \times \left( \matrix{  \frac{1}{|\Pi_{ud}|} & \frac{1}{|\Pi_{us}|} & \frac{1}{|\Pi_{ub}|} \cr
            \frac{1}{|\Pi_{cd}|} &  \frac{1}{|\Pi_{cs}|} &  \frac{1}{|\Pi_{cb}|} \cr
             \frac{1}{|\Pi_{td}|} &  \frac{1}{|\Pi_{ts}|} &  \frac{1}{|\Pi_{tb}|} } \right)
\label{SinPhi=J/Pi}
\end{equation}
where the trigonometric function ${\rm Sin}$ 
must be understood to act independently
on the individual matrix entries as shown.
From Eq.~\ref{SinPhi=J/Pi} the entries in the ${\rm Sin} \, \Phi$ matrix
are clearly inversely proportional to the moduli of the plaquette products, $|\Pi_{\alpha i}|$,
which are themselves each expressible as a product
of four mixing-matrix moduli 
via Eq.~\ref{plaquette}.

Starting from the ${\rm Sin} \, \Phi$ matrix (Eq.~\ref{SinPhi=J/Pi})
and keeping the same (cyclic) definitions of the flavour indices as in Eq,~\ref{plaquette},
we may now define certain products of sines, $\Xi_{\alpha i}$:
\begin{equation}
\Xi_{\alpha_i}:=\sin \Phi_{\alpha j} \sin \Phi_{\alpha k} \sin \Phi_{\beta i} \sin \Phi_{\gamma i}
\label{Xi=ssss}
\end{equation}
mutiplying together
the (four) ${\rm Sin} \, \Phi$ entries  
in the same row and column as $\sin \Phi_{\alpha i}$,
excluding $\sin \Phi_{\alpha i}$ itself. 
Clearly every mixing modulus-squared {\em except} $|V_{\alpha i}|^2$ 
enters in the denominator of the product Eq.~\ref{Xi=ssss},
whereby the $\Xi_{\alpha i}$ 
must be proportional to $|V_{\alpha i}|^2$.
The relevant normalising factor 
may be obtained by summing over any row or column 
(or indeed over both rows and columns).
It should now be clear that:
\begin{eqnarray}
|V_{\alpha i}|^2 & =   & \Xi_{\alpha i} \, / \, (\sum_{\beta}\Xi_{\beta i})
=\Xi_{\alpha i} \, / \, (\sum_{j}\Xi_{\alpha j}) \label{modv21} \\
                 & =   & 3\,\,\Xi_{\alpha i} \, / \, (\sum_{\beta j}\Xi_{\beta j}).
\label{modv22}
\end{eqnarray}
The equivalence of the $\Phi$-matrix and the $(|V|)$-matrix is clearly established,
taking the positive square-root 
(In Eq.~\ref{modv21}-\ref{modv22}, both numerator and denominator are positive). 

In a similar vein we may obtain the
magnitude of the $CP$-invariant $J$: 
\begin{equation}
|J|=9(\prod_{\alpha i}\Xi_{\alpha i})^{1/4} \, / \, (\sum_{\alpha i}\Xi_{\alpha i})^2.
\label{modj}
\end{equation}
Of course in the case of the quarks, 
the ``sense'' of the unitarity triangles 
has already been determined experimentally, 
so that $J$ is anyway already known to be positive for the quarks.
While clearly the $\Xi$-matrix above carries no sign information,
more generally
the $\Phi$-matrix itself (and the ${\rm Sin} \Phi$ matrix) carries explicitly the sign of $J$:
\begin{equation}
J=9\prod_{\alpha i}\sin\Phi_{\alpha i} \; / \; (\sum_{\alpha i}\Xi_{\alpha i})^2
\label{J=9Pisin}
\end{equation}
and the equivalence of $\Phi$ to the complex mixing matrix $V$ is seen to be complete. 
We will return to this point again later in connection with mixing in the lepton sector. 

We may further remark that the normalisation factor in Eq.~\ref{modv21}-\ref{modv22}
is itself a non-trivial flavour-symmetric observable \cite{abcd} \cite{panic}:
\begin{equation}
\sum_{\alpha i}\Xi_{\alpha i}=3 \; / \;  (\prod_{\alpha i} |V_{\alpha i}|^2)
\label{SumXij}
\end{equation}
recognisable as (the reciprocal of) the product of all the mixing-moduli squared \cite{panic}. \\

\noindent {\bf 3. The Sin{\boldmath $\, \Phi$}
              and Cos{\boldmath $\, \Phi$} Matrices in the Wolfenstein Parameterisation.}
\vspace{3mm}

\noindent
Underlying the famous Wolfenstein parametrisation 
\cite{pdg} \cite{wolf} \cite{buras} 
of the CKM matrix,
is the apparent power hierarchy of quark mixing angles 
first remarked upon by Wolfenstein \cite{wolf},
by which $\theta_{12} : \theta_{23} : \theta_{13} \simeq \lambda : \lambda^2 : \lambda^3$,
where $\lambda =\sin \theta_C \sim 0.22$, 
with $\theta_C \simeq \theta_{12}$ the Cabbibo angle \cite{ckm}.
The corresponding hierarchy of $\Phi$-matrix elements 
then follows immediately,
recalling our ``complementary labelling'' of the triangles and angles,
and taking the inner (dot) products in pairs of the rows or columns of $V$:
\begin{eqnarray}
    \matrix{  \hspace{0.7cm} d \hspace{0.05cm}
               & \hspace{0.1cm} s \hspace{0.1cm}
               & \hspace{0.1cm} b  \hspace{0.1cm} }
                                      \hspace{1.5cm}
\hspace{2.0cm}
    \matrix{  \hspace{0.15cm} d \hspace{0.05cm}
               & \hspace{0.05cm} s \hspace{0.05cm}
               & \hspace{0.1cm} b  \hspace{0.05cm} }
                                      \hspace{1.2cm} \nonumber \\\
|V| \sim
\matrix{ u \hspace{0.01cm} \cr
         c \hspace{0.01cm} \cr
         t \hspace{0.01cm} }
\left(\matrix{
1 & \lambda &  \lambda^3 \cr
\lambda & 1 & \lambda^2 \cr
\lambda^3 & \lambda^2   & 1 \cr
} \right)
\hspace{0.35cm}
\Longrightarrow
\hspace{0.35cm}
\Phi \sim
\matrix{ u \hspace{0.01cm} \cr
         c \hspace{0.01cm} \cr
         t \hspace{0.01cm} }
\left(\matrix{
\lambda^2 & 1 & 1 \vspace{5pt} \cr
1 & 1 & 1 \vspace{4pt} \cr
1 & 1 & \lambda^4 \cr
} \right)
 \hspace{0.80cm}
\label{angmat}
\end{eqnarray}
where we have worked initially here only with the bare powers of $\lambda$, for simplicity.
Notice that the centermost row and column of the $\Phi$-matrix, 
corresponding to the \makebox{$(t,u):=c$} and $(b,d):=s$ triangles respectively,
have all angles ``large'' as a result of $\theta_{12}\theta_{23}/\theta_{13} \sim 1$ (hence Figure~1). 
Corresponding to the first row and column of Eq.~\ref{angmat},
the $u$ and $d$ triangles have one small angle 
$\Phi_{ud} \sim \theta_{12}\theta_{13}/\theta_{23} \sim \lambda \lambda^3/\lambda^2 \sim \lambda^2$,
while from the last row and column,
the $t$ and $b$ triangles
have one (very) small angle 
$\Phi_{tb} \sim \theta_{23}\theta_{13}/\theta_{12} \sim \lambda^5/\lambda \sim \lambda^4$
(these various ``long and thin'' triangles are not shown).

Now using the full Wolfenstein parameterisation \cite{wolf} 
in terms of improved paremeters $\bar{\rho}$ and $\bar{\eta}$ \cite{pdg} \cite{buras}, 
we have,
to lowest order (element-wise) in small quantities:
%
\begin{eqnarray}
{{\rm Sin} \, \Phi} \simeq \bar{\eta} \left( \matrix	
{\lambda^2 
         &  b 
               & b 
\cr g 
         &  bg 
               & b 
\cr g 
         &  g 
               & A^2\lambda^4 
} \right)
\label{sinPhi=wolf}
\end{eqnarray}
where
\begin{eqnarray}
b:=\frac{1}{\left[(1-\bar{\rho})^2+\bar{\eta}^2\right]^{\frac{1}{2}}},\quad
g:=\frac{1}{(\bar{\rho}^2+\bar{\eta}^2)^{\frac{1}{2}}}.
\label{define B ang g}
\end{eqnarray}
Notice that the 
${\rm Sin} \, \Phi$ matrix is proportional 
to the (signed) $CP$-violating quantity $\bar{\eta}$.

We may now use Eq.~\ref{sinPhi=wolf} together with 
Eqs.~\ref{Xi=ssss} and Eq.~\ref{J=9Pisin} above,
to recover the usual Wolfenstein approximation for $J$:
\begin{equation}
J=\frac{\prod_{\alpha i} \sin\Phi_{\alpha i}}{(\sum_{\alpha}\Xi_{\alpha i})^2}
\simeq \frac{\bar{\eta}^9A^2\lambda^6b^4g^4}{(\bar{\eta}^4b^2g^2)^2} 
\simeq \bar{\eta} A^2\lambda^6.
\label{elementJmatrix3}
\end{equation}

The $ {\rm Cos} \; \Phi$ matrix similarly takes the form:
\begin{eqnarray}
{\rm Cos} \, \Phi \simeq \left( \matrix{	1 & b(1-\bar{\rho}) & -b(1-\bar{\rho}) \cr
			g\bar{\rho} & bg[\bar{\eta}^2-\bar{\rho}(1-\bar{\rho})] & b(1-\bar{\rho}) \cr
			-g\bar{\rho} & g\bar{\rho} & 1 } \right).
\label{CosPhi=wolf}
\end{eqnarray}
A vanishing ${\rm Cos}\; \Phi$ element
implies a right angle in the $\Phi$ matrix,
e.g. $\cos\Phi_{cs}=0 \Rightarrow \Phi_{cs}=90^o$ 
\cite{panic} \cite{xing09} (see Eq.~\ref{Phi=degrees}), 
corresponding to the exact constraint:
\begin{equation}
\bar{\eta}^2-\bar{\rho}(1-\bar{\rho})=0 \hspace{0.8cm} {\rm ie.} \hspace{0.3cm} \bar{\rho}=\bar{\rho}^2+\bar{\eta}^2.
\end{equation}
We may remark that, 
while the ${\rm Sin} \; \Phi$ matrix 
is clearly independent 
of a possible choice to work with 
external rather than internal angles,
the ${\rm Cos}\; \Phi$ matrix would change sign.

The $K$-matrix is just $J$ times the ${\rm Cot} \, \Phi$ matrix:
\begin{eqnarray}
K=J\,{\rm Cot}\,\Phi \simeq \frac{J}{\bar{\eta}} \left(\matrix {1/\lambda^2 & (1-\bar{\rho}) & -(1-\bar{\rho}) \cr
					  \bar{\rho} & \bar{\eta}^2-\bar{\rho}(1-\bar{\rho}) & (1-\bar{\rho}) \cr
					 -\bar{\rho} & \bar{\rho} & 1/A^2\lambda^4} \right).
\label{Kmatrix}
\end{eqnarray}
In a previous publication \cite{abcd} (and see \cite{panic})
we have considered the possibility that it is 
${\rm Det} \, K$ 
which vanishes exactly
(rather than $\cos \Phi_{cs}$ as above).
Eq.~\ref{Kmatrix} readily gives:
\begin{equation} 
{\rm Det}K\simeq (J/\bar{\eta})^3(1/A^2\lambda^6)\{
[\bar{\eta}^2-\bar{\rho}(1-\bar{\rho})]
-\lambda^2 \bar{\rho}(1-\bar{\rho}) \}
\end{equation}
which is valid for $|\bar{\eta}^2-\bar{\rho}(1-\bar{\rho})|$ $\simlt$ ${\cal O}(\lambda)$.
Setting ${\rm Det} \, K$ exactly to zero then predicts:
\begin{equation}
\cos \Phi_{cs} = bg[\bar{\eta}^2-\bar{\rho}(1-\bar{\rho})] \simeq
\lambda^2 b g \bar{\rho}(1-\bar{\rho}) 
\simeq \bar{\eta}\lambda^2,
\end{equation}
i.e. we predict a small correction to $\cos\Phi_{cs}= 0$ above, 
such that $\cos\Phi_{cs} \simeq \bar{\eta}\lambda^2$ \cite{panic}\cite{abcd}.

Concerning RGE evolution,
it should be remarked that
several authors \cite{evol} have noted that
of the four Wolfenstein parameters 
($\lambda$, $A$, $\bar{\rho}$, $\bar{\eta}$)
fixing the CKM matrix, 
it is only the parameter $A$ which evolves
to leading order in the SM and MSSM \cite{evol}.
We then have from Eq.~\ref{sinPhi=wolf}
that, while mixing moduli $|V_{\alpha i}|$
and UT sides etc.\
definitely should evolve,
by contrast in the angle matrix $\Phi$ 
only the very small bottom-right corner element 
$\Phi_{tb} \sim A^2\lambda^4$ 
should evolve significantly,
so that
the angle matrix $\Phi$,
at least as regards its general appearance (Eq.~\ref{Phi=degrees})
may be said to be largely invariant.
We do not consider 
RGE evolution any further here, 
only remarking that 
the angle matrix $\Phi$ (Eq.~\ref{Phi=degrees})
may already be revealing its basic form
at the very highest energy scales. \\

\noindent {\bf 4. Prior Nomenclatures, Current Results and some Future Prospects}
\vspace{3mm}

\noindent 
As a prelude to discussing the experimental measurements, errors etc.\
on the angle-matrix entries,
it will be necessary to establish the correspondence 
with some of the historical 
namings of UT angles,
at least in those cases for which prior namings exist.  
In particular, UT angles $\beta$, $\alpha$, $\gamma$, 
also known as $\phi_1$, $\phi_2$, $\phi_3$,
have of course already been intensively studied theoretically and experimentally
(the ``switch'' in ordering between the Babar \cite{babar01} $\alpha, \beta, \gamma$ 
and Belle \cite{belle01} $\phi_1, \phi_2, \phi_3$ nomenclatures 
is well-known and 
perhaps a little unfortunate, 
but can hardly be a cause of any major confusion).
From our definitions
\footnote{
We note that UT angles
in the literature \cite{pdg} e.g.\ $\alpha$, $\beta$, $\gamma$,
are often defined in terms of {\em ratios} of mixing matrix elements,
e.g.\ $\alpha=\arg (-V_{td}V_{tb}^*/V_{ud}V_{ub}^*)$, 
$\beta=\arg (-V_{cd}V_{cb}^*/V_{td}V_{tb}^*)$ etc.\ \cite{pdg}.
The definition of Section~1-2 (Eq.~\ref{Phi=ArgPi}-\ref{Pi}) 
with all four relevant mixing elements multiplied 
systematically on an essentially equal footing, 
is of course entirely equivalent 
to those definitions in terms of ratios~\cite{pdg}.}
(Eqs.~\ref{Phi=ArgPi}-\ref{Pi})
we have $\Phi_{us}=\beta=\phi_1$, $\Phi_{cs}=\alpha=\phi_2$ and $\Phi_{ts}=\gamma=\phi_3$,
comprising the central column of our full $\Phi$-matrix: 
\begin{eqnarray}
    \matrix{  \hspace{1.2cm} d \hspace{1.2cm}
               & \hspace{1.2cm} s \hspace{1.2cm}
               & \hspace{1.2cm} b  \hspace{1.5cm} }
\hspace{0.6cm} \nonumber \\
\Phi=
\matrix{ u \hspace{0.01cm} \cr
         c \hspace{0.01cm} \cr
         t \hspace{0.01cm} }
\left(\matrix{
\Phi_{ud}=\beta_s =\chi& \Phi_{us} =\beta =\phi_1 & \Phi_{ub} \vspace{5pt} \cr
\Phi_{cd} =\gamma' =\gamma-\delta \gamma & \Phi_{cs} =\alpha =\phi_2 & \Phi_{cb} =\beta+\delta \gamma \vspace{4pt} \cr
\Phi_{td} & \Phi_{ts} =\gamma =\phi_3 & \Phi_{tb} = \beta_K = \chi' \cr
} \right).
\hspace{0.35cm}
\label{nota}
\end{eqnarray}

In the $B_s$-sector, 
the angle $\Phi_{ud}$ (top-left entry in Eq.~\ref{nota})
is quite naturally 
seen as the analogous angle to $\Phi_{us}=\beta$ in the $B_d$-sector,
whereby 
one has $\Phi_{ud}=\beta_s$ \cite {d0}. 
Note however that since the direct measurements in the $B_s$-sector 
are potentially very sensitive to possible new physics contributions,
the more specific designation $\beta_s^{SM}$ is also sometimes used \cite{cdf}
to indicate explicitly the SM contribution alone, 
which ultimately defines our $\Phi$-matrix here.
Since $\Phi_{ud}$ is often denoted $\chi$ in the theory literature \cite{stone}
we take $\Phi_{ud}=\beta_s=\beta_s^{SM}=\chi$. 
Other notations 
have sometimes been used (e.g.\ $-\phi_s/2$ \cite{d0})
especially to denote the directly measured empirical angle 
inclusive of any new physics
(this distinction is far from academic \cite{cdf} \cite{d0} as discussed below). 
 

To complete a set of four independent angles
(clearly $\alpha$, $\beta$, $\gamma$ above are {\em not} mutually independent)
one might easily take $\Phi_{cd}$, 
to complete a ``$\Phi$-plaquette'' \cite{pkq}
(with $\Phi_{us}$, $\Phi_{cs}$ and $\Phi_{ud}$)
enabling all other angles to be readily calculated,
summing rows and columns to $180^o$.
We note that $\Phi_{cd}$ has been denoted
$\gamma' := \gamma-\delta \gamma$ \cite{lhcb}.
This latter notation 
exploits the fact that the row-based 
$c$-triangle and the column-based $s$-triangle,
being otherwise independent,
have the angle $\alpha$ in common (see again Figure~1).
We may thus write $\Phi_{cd}=\gamma-\delta \gamma$
and (equivalently) 
also $\Phi_{cb}=\beta+\delta \gamma$ 
(since $\alpha+\beta+\gamma=180^o$). 
As we will see, however,
neither $\Phi_{cd}$, $\Phi_{cb}$ nor $\delta \gamma$
is necessarily an optimal choice 
to complete the set in practice.
Indeed, we will rather take $\Phi_{tb}$ 
(bottom-right entry in Eq.~\ref{nota}) 
as our fourth parameter here \cite{lebed}, 
as will be discussed in more detail below.

Turning now to the experimental results themselves, 
as is well-known,
a number of direct measurements of the angle $\Phi_{us}=\beta$ 
in the $B_d \rightarrow J/\psi K_s$ mode  
have been carried out over many years, 
especially at the $B$-factories \cite{babarbeta} \cite{bellebeta} 
yielding a world average $\sin 2 \beta =0.684\pm 0.022$ \cite{moriond09}.
Ambiguities can be resolved \cite{cos2beta} 
to give finally: $\Phi_{us}=\beta=21.58^o \pm 0.86^o$ \cite{moriond09}.
The combined decay modes
$B_d \rightarrow \pi \pi$, $B_d \rightarrow \rho \rho$ 
and $B_d \rightarrow \rho \pi$ yield: 
$\Phi_{cs}=\alpha \simeq 90.6^o \pm 4.0^o$ \cite{babaralpha} \cite{bellealpha} \cite{moriond09}
(consistent with $\alpha=90^o$ as earlier noted \cite{panic}). 
Direct measurements of the angle $\gamma$ 
yield $\Phi_{ts} =\gamma \sim 70^o \pm 30^o$ \cite{babargamma} \cite{bellegamma} \cite{moriond09}
consistent with SM unitarity ($\alpha+\beta+\gamma=180^o$).
LHCb \cite{lhcb}, 
followed by Super Flavour Factories~\cite{superb},
can ultimately reduce the errors
on the angles $\Phi_{cs}=\alpha$ and $\Phi_{us}=\beta$ 
by as much as a factor of four or so,
with both projects vastly improving the measurement 
of $\Phi_{ts}=\gamma$ 
to reach an error of perhaps $\pm 2^o$ or better.

The angle 
$\Phi_{ud}=\beta_s=\chi$ 
is in fact already highly constrained 
by existing indirect measurements through SM unitarity constraints,
giving $\Phi_{ud} 
\simeq \bar{\eta} \lambda^2(1+(1-\bar{\rho})\lambda^2) \sim 1.04^o \pm 0.05^o$ 
using latest fits \cite{moriond09}.
Direct measurements of $\Phi_{ud}=\beta_s$
in the $B_s^0 \rightarrow J/\psi \phi$ channel
are potentially sensitive to new physics contributions
and indeed, 
combining recent CDF-II \cite{cdf} and D0 \cite{d0} results 
gives $\Phi_{ud} = \beta_s \simeq (19 \pm 7)^o$,
approximately 2.6$\sigma$ from the expected SM value above.
LHCb \cite{lhcb} is expected to make 
the definitive measurent of $\Phi_{ud}=\beta_s$, 
ultimately to better than $\pm 1^o$ accuracy.
Such measurements, if anomalies persist,
may eventually need to be re-interpreted,
possibly with reference to a new/different angle matrix 
(or even several such matrices) relevant to the new physics,
after subtraction of the SM contribution above.

Similarly, our best knowledge of 
$\Phi_{tb} \simeq \bar{\eta} A^2 \lambda^4(1+\lambda^2) 
\sim 0.035^o \pm \raisebox{0.9ex}{$0.003^o$} \raisebox{-0.9ex} {\hspace{-11.5mm}$0.002^o$}$ \cite{moriond09}
comes indirectly through SM unitarity constraints.
$\Phi_{tb}$ is sometimes denoted $\beta_K$ \cite{babarphys} 
being at least in principle related to $CP$ violation in $K^0$ mixing
(and similarly in $D^0$ mixing).
In the theory literature
$\Phi_{tb}$ is sometimes denoted $\chi'$ \cite{stone}.
One chooses $\Phi_{tb} =\beta_K =\chi'$ 
as the fourth primary parameter here,
rather than $\Phi_{cd}=\gamma-\delta \gamma$ above,
since $\delta \gamma$ and $\beta_s$ are 
anyway very nearly equal
\footnote{Indeed one occasionally sees the $\bar{c}$-triangle of Figue~1b
with its two base angles labelled $\gamma-\chi$ and $\beta+\chi$ \cite{muheim}
or (equivalently) $\gamma-\beta_s$ and $\beta+\beta_s$ \cite{golutvin} respectively.
This is a good working approximation in the SM
since $\Phi_{ud}=\beta_s=\chi \simeq 1.04^o$ (Eq.~\ref{errors})
while $\delta \gamma =\Phi_{ud}-\Phi_{tb} \simeq 1.00^o$ (Eq.~\ref{errors})
i.e. currently indistinguishable from each other within the errors 
on the direct measurement of $\Phi_{ud}=\beta_s=\chi$.},
due to the smallness of $\Phi_{tb}$
(note that $\Phi_{tb}=\beta_s-\delta \gamma$). 

Given $\Phi_{us}$, $\Phi_{cs}$, $\Phi_{ud}$ and $\Phi_{tb}$ above, 
all the remaining entries in the $\Phi$-matrix 
can now be determined in an obvious way,
summing rows and columns to $180^o$. We find: 
\begin{eqnarray}
    \matrix{  \hspace{1.0cm} d \hspace{1.2cm}
               & \hspace{1.2cm} s \hspace{1.3cm}
               & \hspace{1.2cm} b  \hspace{1.1cm} }
                                      \hspace{1.5cm} \nonumber \\\
\Phi \simeq
\matrix{ u \hspace{0.01cm} \cr
         c \hspace{0.01cm} \cr
         t \hspace{0.01cm} }
\left(\matrix{
1.04^o \pm 0.05^o & 21.58^o\pm 0.86^o &  157.38^o \mp 0.89^o \cr
66.82^o \mp 4.20^o & 90.60^o \pm 4.00^o & 22.58^o \pm 0.89^o \cr
112.14^o \pm 4.21^o & 67.82^o \mp 4.22^o  & 0.035^o \pm 0.003^o \cr
} \right), \hspace{0.80cm}
\label{errors}
\end{eqnarray}
where the main correlations are indicated by the signs on the errors. 
In Eq.~\ref{errors} 
we have padded the value of
$\Phi_{cs}$ given above by an extra decimal digit,
$\Phi_{cs} \rightarrow 90.60^o \pm 4.00^o$, 
for uniformity with the higher precision of the other input angles. 
Also, the asymmetric error on $\Phi_{tb}$ quoted above
has been (conservatively) symmetrised,
$\Phi_{tb} \rightarrow 0.035^o \pm 0.003^o$, 
for simplicity of presentation in the matrix (Eq.~\ref{errors}).

In the leptonic case the angles may be 
determined in neutrino oscillation experiments.
The measurement of at least one non-zero $CP$-violating asymmetry 
in the leptonic case will be needed  
to determine the ``sense'' of the leptonic triangles 
(as for the quarks, {\em triangles} are defined
with the inner-product arguments in ascending (cyclic) mass order,
and the ``sense'' similarly from the mass-ordering of the sides).  
The measurement 
will 
determine the sign of the leptonic $J$ 
and hence the (common) sign of all the entries in the angle matrix, so that
row and column sums in the leptonic case may sum to $+180^o$ as for the quarks,
or possibly to $-180^o$, as remains to be seen. \\

\noindent {\bf 5. Concluding Remarks}
\vspace{3mm}

\noindent In this paper we have introduced and developed 
the concept of a matrix of unitarity triangle angles for the quarks, 
as for the leptons,
showing that it carries equivalent information to the complex mixing matrix itself,
with the added advantage of being basis- and phase-convention independent and hence fully observable.
Each row and each column of the angle matrix 
lists directly the angles of a specific unitarity triangle
(three row-based and three column-based triangles,
making six unitarity triangles in all, 
of which four have largely RGE-invariant shape in the SM and MSSM).
Individual angles are labelled in 
a systematic, well-defined, physically-motivated way.
The marked hierarchy of the CKM elements, 
as reflected in the Wofenstein parameterisation,
translates in the angle matrix
into just two small angles: 
$\Phi_{ud} \sim \lambda^2$ and $\Phi_{tb} \sim \lambda^4$.
The centremost row and column 
of the angle matrix each comprise
all ``large'' angles and correspond to the 
two most familiar unitarity triangles, 
which have been long studied theoretically and experimentally
already in the context of $B \leftrightarrow \bar{B}$ oscillations. \\

\noindent {\bf Acknowledgement}
\vspace{1mm}

\noindent
This work was supported by the UK Science and Technology Facilities Council (STFC).
PFH acknowledges the hospitality 
of the Centre for Fundamental Physics (CfFP)
Rutherford Appleton Laboratory.
We also acknowledge useful contributions at various points
from \mbox{T.\ Gershon}, \mbox{F.\ Long}, D. Roythorne, S.\ Sun, M.\ Williams and F.\ Wilson.
\vspace{-3mm}



\begin{thebibliography}{99}  


\bibitem{bigi} A. B. Carter and A.I. Sanda, Phys. Rev. D23 (1981) 1567. \\
I.I. Bigi and A.I. Sanda, Nucl. Phys. B193 (1981) 85.


\bibitem{belle} 
``Experimental Review of CP Violation and CKM Angles at B Factories'', \\
K.\ F.\ Chen at DISCRETE'08,
Dec.\ 11-16, 2008, IFIC, Valencia, Spain. \\
(Also avaiable at: http://belle.kek.jp/belle/talks/DIS08/Chen.pdf ). 


\bibitem{ckm} N. Cabibbo, Phys.\ Rev.\ Lett.\ 10 (1963) 531. \\
M.\ Kobayashi and T.\ Maskawa, Prog.\ Thor.\ Phys.\ 49 (1973) 652. 


\bibitem{aleksan} R. Aleksan, B. Kayser and D. London, \\
Phys.\ Rev.\ Lett.\ 73 (1994) 18. hep-ph/9403341


\bibitem{lebed} R.\ F.\ Lebed, Phys.\ Rev.\ D 55 (1997) 348. arXiv:hep-ph/9607305 \\
G.\ C.\ Branco, L.\ Lavoura and J.\ P.\ Silva, ``$CP$ Violation'', OUP (1999).



\bibitem{bhs} J.\ D.\ Bjorken, P.\ F.\ Harrison and W.\ G.\ Scott, \\
Phys.\ Rev.\ D 74 (2006) 073012. hep-ph/0511201


\bibitem{wolf} L.\ Wolfenstein, Phys.\ Rev.\ Lett.\ 51 (1983) 1945. 


\bibitem{branco} 
F.\ J.\ Botella and Ling-Lie Chau, Phys.\ Lett.\ B168 (1986) 97. \\
C. Jarlskog, Proc. IVth LEAR Workshop,
Villar-sur-Ollon, Switzerland (1987). \\ 
G.\ C.\ Branco and L. Lavoura, Phys. Lett. B 208 (1988) 123. 
C.\ Hamzaoui, Phys.\ Rev.\ Lett.\ 61 (1988) 35. 
C.\ Jarlskog and R. Stora, Phys. Lett. B 208 (1988) 268. 
G.\ Auberson, Phys. Lett. B 216 (1989) 167. 
L.\ Lavoura, Phys.\ Lett.\ B 223 (1989) 97. 
J.\ D.\ Bjorken, Phys.\ Rev.\ D 39 (1989) 1396.

\bibitem{pdg} C.\ Amsler et al.\ (Particle Data Group)
Phys.\ Lett.\ B 667 (2008) 145. \\
Also available at: http://pdg.lbl.gov/2008/reviews/rpp2008-rev-ckm-matrix.pdf



\bibitem{plaq} J.\ D.\ Bjorken and I. Dunietz Phys.\ Rev.\ D 36 (1987) 2109. \\
C.\ Jarlskog, Phys.\ Rev.\ Lett.\ 55 (1985) 1039. \\
I.\ Dunietz, O.\ W.\ Greenberg and D.\ Wu, Phys.\ Rev.\ Lett.\ 55 (1985) 2935 \\
D.\ Wu Phys.\ Rev.\ D 33 (1986) 860. \\
D.\ Du, I.\ Dunietz and D.\ Wu Phys.\ Rev.\ D 34 (1986) 3414. \\
J.\ F.\ Nieves and P.\ B.\ Pal Phys.\ Rev.\ D 36 (1987) 315.


\bibitem{ven09} P.\ F.\ Harrison, W.\ G.\ Scott,
presented by W.\ G.\ Scott at Neutrino Telescopes, 
Venice, Italy (2009).  arXiv:0906.3077 [hep-ph].


\bibitem{pkq} P.\ F.\ Harrison, W.\ G.\ Scott and T.\ J.\ Weiler,
Phys.\ Lett.\ B 641 (2006) 372. hep-ph/0607335.


\bibitem{weil} D.\ J.\ Wagner and T.\ J.\ Weiler, Phys.Rev. D59 (1999) 113007. hep-ph/9801327. \\
T.\ J.\ Weiler and D.\ J.\ Wagner,
Proceedings of the 2nd Latin American Symposium 
on High Energy Physics, 
San Juan, Puerto Rico (1998) hep-ph/9806490. 


\bibitem{loops} B.\ Durhuus and J. M. Leinaas, Physica Scripta 25 (1982) 504. CERN-TH-3110 \\
H-M.\ Chan, S.\ T.\ Tsou, Int.\ J.\ Mod.\ Phys.\ A14 (1999) 2139. hep-th/9904102 


\bibitem{cecj} 
C.\ Jarlskog, Z.\ Phys.\ C 29 (1985) 491. \\
C. Jarlskog, Phys. Rev. D35 (1987) 1685; 
{\em ibid} D36 (1987) 2128. \\
C. Jarlskog, ``$CP$ violation'', Ed.\ C.\ Jarlskog (World Scientific 1989).

\bibitem{abcd} 
P.\ F.\ Harrison, D.\ R.\ J.\ Roythorne and W.\ G.\ Scott,
Phys.\ Lett.\ B 657 (2007) 210. arXiv:0709.1439 [hep-ph]. \\
P.F. Harrison, D.R.J. Roythorne, W.G. Scott,  
Proc.\ 43rd Rencontres de Moriond, 
EW Interactions, La Thuile, Italy (2008) arXiv:0805.3440 [hep-ph].


\bibitem{panic} 
P.\ F.\ Harrison, D.\ R.\ J.\ Roythorne and W.\ G.\ Scott,
Proc.\ 18th Particle and Nuclei Intl.\ Conf.\ (PANIC08),
Eilat, Israel (2008).  arXiv:0904.3014 [hep-ph].





\bibitem{buras} A.\ J.\ Buras, Phys.\ Rev.\ D 50 (1994) 3433. hep-ph/9403384


\bibitem{xing09} 
H.\ Fritzsch and Z.-z.\ Xing, Phys.\ Lett.\ B 353 (1995) 114. \hspace{2mm}  hep-ph/9502297. \\
I.\ Masina and C.\ A.\ Savoy, Nucl.\ Phys.\ B755 (2006) 1. hep-ph/0603101; \\
Phys.\ Lett.\ B642 (2006) 472. hep-ph/0606097.
\hspace{2mm} Z.-z.\ Xing, arXiv:0904.3172. 




\bibitem{evol}
S.\ R.\ Juarez et al., Phys.\ Rev.\ D66 (2002) 116007. hep-ph/0206243 \\
B.\ A.\ Kniehl et al. 
Phys.\ Rev.\ D62 (2002) 073010. hep-ph/0005060

\bibitem{babar01} B. Aubert et al. (BaBar collab.), 
Phys.\ Rev.\ Lett.\  86 (2001) 2515. \\
J.D. Bjorken SLAC-PUB-5389, 
18th SLAC Summer Inst. (1990) 0167-198.

\bibitem{belle01} A. Abashian et al. (Belle collab.), 
Phys.\ Rev.\ Lett.\  86 (2001) 2509. \\
I.\ Bigi et al.\ 
in $CP$ Violation, ed C. Jarlskog, World Scientific, Singapore (1989) 175-248.
SLAC-PUB-4476, RU-DOE/ER/40352-24 (1987).


\bibitem{cdf} T.\ Aaltonen et al. (CDF collab.)
Phys.\ Rev.\ Lett.\ 100 (2008) 161802 \\
arXiv: 0712.2397 [hep-ph].
see also CDF/ANAL/BOTTOM/PUBLIC/9458.

\bibitem{d0} V.\ M.\ Abazov et al. (D0 collab.)
Phys.\ Rev.\ Lett.\ 101 (2008) 241801 \\
arXiv: 0802.2255 [hep-ph]


\bibitem{stone}
F. J. Botella et al. 
Nucl.\ Phys.\ B, 651 (2003) 174. \\
J.A. Aguilar-Saavedraa et al.\  
Nucl.\ Phys.\ B 706 (2005) 204. \\
M.\ Artuso, E.\ Barberio, S.\ Stone.
PMC Physics A (2009) 3:3. arXiv:0902.3743







\bibitem{lhcb}
S. Amato et al. LHCb Technical proposal, CERN LHCC 98-4 LHCC/P4 (1998). 



\bibitem{babarbeta} B. Aubert et al. (BaBar collab.), 
Phys.\ Rev.\ Lett.\ 87 (2001) 091801. 
Phys.\ Rev.\ D 66 (2001) 032003.
Phys.\ Rev. Lett.\ 89 (2002) 201802. 
Phys.\ Rev. Lett.\ 94 (2005) 161803. 
Phys.\ Rev.\ Lett.\ 99 (2007) 171803. 
B. Aubert et al. (Babar collab.) 
Phys.~Rev.~D. 79 (2009) 072009, arXiv:0902.1708 [hep-ex]. 


\bibitem{bellebeta}
K. Abe et al. (Belle collab.) Phys. Rev. Lett. 87 (2001) 091802. \\ 
K. Abe et al., Phys. Rev. D 66 (2002) 032007; 
Phys. Rev. D 66 (2002) 071102(R). 
K. Abe et al. Phys. Rev. D 71 (2005) 072003. \\ 
K-F. Chen et al. (Belle Collab.) Phys. Rev. Lett. 98 (2007) 031802. 


\bibitem{cos2beta}
I.\ Dunietz et al. Phys.\ Rev.\ D 43 (1991) 2193. \\
J.\ Charles et al. Phys.\ Rev.\ D 58 (1998) 114021. \\ 
B.\ Aubert et al. (Babar Collab.) Phys.\ Rev.\ D 71 (2005) 032005. hep-ex/0411016. \\ 
R.\ Itoh et al. (Belle collab.) Phys.\ Rev.\ Lett.\ 95 (2005) 091601. hep-ex/0504030. \\ 
J.\ Charles et al. Phys.\ lett.\ B 425 (1998) 375. hep-ph/9801363. \\
T.\ Latham and T.\ Gershon, J.\ Phys.\ G36 (2009) 025006. arXiv:0809.0872


\bibitem{moriond09}
http://ckmfitter.in2p3.fr/plots\underline{ }Moriond09/ckmEval\underline{ }results.html \\
J.\ Charles et al. (CKMfitter Group) Eur.\ Phys.\ J.\ C 41 (2005) 1. hep-ph/0406184. \\
http://www.slac.stanford.edu/xorg/hfag/triangle/summer2008/index.shtml \\
E.\ Barberio et al. (HFAG collab.) arXiv:0808.1297. \\
M. Bona et al. (UTfit Collab.) JHEP 0507 (2005) 028. hep-ph/0501199.



\bibitem{babaralpha}
B. Aubert et al. (Babar Collab.) Phys.\ Rev.\ Lett.\ 94 (2005) 181802. 
Phys.\ Rev. Lett.\ 94 (2005) 131801. 
Phys.\ Rev.\ Lett.\ 95 (2005) 151803. \\ 
B. Aubert et al. (Babar collab.) SLAC-PUB-13452 BABAR-PUB-08/053 
arXiv: 0901.3522 [hep-ex] submitted to Phys. Rev. Lett. 



\bibitem{bellealpha}
K. Abe et al. (Belle collab) Phys. Rev. Lett. 93 (2004) 021601. \\ 
C. C. Wang et al. (Belle collab.) PRL 94 (2005) 121801.  \\ 
A. Somov, A. J. Schwartz et al. (Belle collab.) Phys. Rev.\ Lett.\ 96 (2006) 171801. 



\bibitem{babargamma}
B. Aubert et al. (Babar collab.) Phys.\ Rev. Lett. 95 (2005) 121802. \\ 
B.\ Aubert et al. (Babar collab.) Phys.\ Rev.\ D78 (2008) 034023. \\ 
T.\ Gershon, Phys.Rev.D79 (2009) 051301. arXiv:0810.2706.
 


\bibitem{bellegamma}
A. Poluektov (Belle collab.)  Phys.\ Rev.\ D 70 (2004) 072003. \\ 
A. Poluektov (Belle collab.) Phys. Rev. D 73 (2006) 112009. \\ 
K. Abe (Belle collab.) BELLE-CONF-0801, arXiv:0803.3375 


\bibitem{superb} 
T. Browder et al.\ JHEP 0802:110, 2008. arXiv:0710.3799 [hep-ph]. \\ 
C-h.\ Cheng, also A.\ Bevan, at ``New Physics with SuperB'', Warwick, April 2009: \\
http://www2.warwick.ac.uk/fac/sci/physics/research/epp/meetings/superb2009. \\
K.\ Abe et al.\ SuperKEKB LoI, KEK Report 04-4. see: http://superb.kek.jp/


\bibitem{babarphys}
The Babar Physics Book, eds. P. F. Harrison and H. R. Quinn (1998).


\bibitem{muheim}
F.\ Muheim ``Flavour in the Era of LHC'' HEP Forum, Abingdon, UK (2007). \\
http://hepwww.rl.ac.uk/accel/forum/2007b/forum.html

\bibitem{golutvin}
A.\ Golutvin ``The Flavour Physics Potential of LHCb'' $B$-Factory Symposium, \\
QMUL, London, UK (2009). http://pprc.qmul.ac.uk/babarSymposium/



\end{thebibliography}
\end{document}